\documentclass[10pt, conference, compsocconf]{IEEEtran}
\usepackage{amsmath}
\usepackage{graphicx}
\begin{document}
%
% paper title
% can use linebreaks \\ within to get better formatting as desired
\title{The shortest time and/or the shortest path strategies \\ in a CA FF pedestrian dynamics model}

% author names and affiliations
% use a multiple column layout for up to two different
% affiliations

% conference papers do not typically use \thanks and this command
% is locked out in conference mode. If really needed, such as for
% the acknowledgment of grants, issue a \IEEEoverridecommandlockouts
% after \documentclass

% for over three affiliations, or if they all won't fit within the width
% of the page, use this alternative format:
%
\author{\IEEEauthorblockN{Ekaterina Kirik\IEEEauthorrefmark{1}\IEEEauthorrefmark{2},
Tat'yana Yurgel'yan\IEEEauthorrefmark{2}, and Dmitriy Krouglov
\IEEEauthorrefmark{1}\IEEEauthorrefmark{3}}
\IEEEauthorblockA{\IEEEauthorrefmark{1}
Institute of Computational Modelling of Siberian Branch of Russian Academy of Sciences,\\
Krasnoyarsk, Akademgorodok, Russia, 660036\\
Email: kirik@icm.krasn.ru}
\IEEEauthorblockA{\IEEEauthorrefmark{2}Siberian Federal University, Institute of Mathematics\\
79 Svobodniy av., Krasnoyarsk, Russia, 660041}
%\\
%Email: homer@thesimpsons.com}
\IEEEauthorblockA{\IEEEauthorrefmark{3}V.N. Sukachev Institute of forest of Siberian Branch of Russian Academy of Sciences,\\
Krasnoyarsk, Akademgorodok, Russia, 660036}
}

%The shortest time and/or the shortest path strategies in a CA FF pedestrian dynamics model
%Ekaterina Kirik, Tat'yana Yurgel'yan, and Dmitriy Krouglov
% 4 figures, 4 pages; the paper was submitted to the Crowds and pedestrian behavior workshop (CROWD) for 2009 IEEE/WIC/ACM International Conference on Web Intelligence and Intelligent Agent Technology, University of Milano Bicocca, Milan
%

% use for special paper notices
%\IEEEspecialpapernotice{(Invited Paper)}

% make the title area
\maketitle

\begin{abstract}
This paper deals with a mathematical model of a pedestrian
movement. A stochastic cellular automata (CA) approach is used
here. The Floor Field (FF) model is a basis model. FF models imply
that virtual people follow the shortest path strategy. But people
are followed by a strategy of the shortest time as well. This
paper is focused on how to mathematically formalize and implement
to a model these features of the pedestrian movement. Some results
of a simulation are presented.

\end{abstract}

\begin{IEEEkeywords}
Cellular automata; pedestrian dynamics; transition probabilities;
artificial intelligence
\end{IEEEkeywords}

% For peer review papers, you can put extra information on the cover
% page as needed:
% \ifCLASSOPTIONpeerreview
% \begin{center} \bfseries EDICS Category: 3-BBND \end{center}
% \fi
%
% For peerreview papers, this IEEEtran command inserts a page break and
% creates the second title. It will be ignored for other modes.
\IEEEpeerreviewmaketitle

\section{Introduction}
% no \IEEEPARstart
A stochastic cellular automata (CA) model of pedestrian flow is
considered here. Our model takes inspiration from stochastic floor
field~(FF) CA model~\cite{ExtFFCAMod} that provides pedestrians
with a map which ``shows'' the shortest distance from current
position to a target. In this paper we focus on mathematical
formalizing and implementation to the model such behavioral
aspects of decision making process as: while moving people follow
at least two strategies
--- the shortest path and the shortest time. Strategies may vary,
cooperate and compete depending on current position.

This is a next attempt to extend basis FF model towards a
behavioral aspect making more flexible/realistic decision making
process and improve simulation of individual and collective
dynamics of people flow.

%\subsection{Subsection Heading Here}
%Subsection text here.

%\subsubsection{Subsubsection Heading Here}
%Subsubsection text here.

\section{Statement of the problem} The space (plane) is known and
sampled into cells $40cm \times 40cm$ which can either be empty or
occupied by one pedestrian (particle) only~\cite{ExtFFCAMod}.
Cells may be occupied by walls and other nonmovable
obstacles. So space is presented by 2 matrixes: %\vspace{-3mm}
$$f_{ij}= \left\{%
\begin{array}{ll}
    1, & \hbox{cell $(i,j)$ is occupied by a pedestrian;} \\
    0, & \hbox{cell $(i,j)$ is empty,} \\
\end{array}%
\right. $$

$$w_{ij}= \left\{%
\begin{array}{ll}
    1, & \hbox{cell $(i,j)$ is occupied by an obstacle;} \\
    0, & \hbox{cell $(i,j)$ is empty.} \\
\end{array}%
\right. $$

A \textit{Static Floor Field} (SFF) $S$ is used in the model.
Field $S$ coincides with the sampled space. A value of each
$S_{i,j}$ saves the shortest distance from cell $(i,j)$ to a
nearest exit; i.e., S increases radially from exit cells where
$S_{i,j}$ are zero. It doesn't evolve with time and isn't changed
by the presence of the particles. One can consider $S$ as a map
that pedestrians use to move to the nearest exit.

\begin{figure}[h]
\begin{center}
\begin{picture}(230,90)
\put(0,0){\line(1,0){90}}%
\put(0,30){\line(1,0){90}}%
\put(0,60){\line(1,0){90}}%
\put(0,90){\line(1,0){90}}%
\put(0,0){\line(0,1){90}}%
\put(30,0){\line(0,1){90}}%
\put(60,0){\line(0,1){90}}%
\put(90,0){\line(0,1){90}}%
\put(45,45){\vector(-1,0){30}}%
\put(45,45){\vector(1,0){30}}%
\put(45,45){\vector(0,1){30}}%
\put(45,45){\vector(0,-1){30}}%
\put(45,45){\circle*{200}}%
\put(140,0){\line(1,0){90}}%
\put(140,30){\line(1,0){90}}%
\put(140,60){\line(1,0){90}}%
\put(140,90){\line(1,0){90}}%
\put(140,0){\line(0,1){90}}%
\put(170,0){\line(0,1){90}}%
\put(200,0){\line(0,1){90}}%
\put(230,0){\line(0,1){90}}%
\put(152.5,9.5){\Large 0}%
\put(211.5,9.5){\Large 0}%
\put(152.5,69.5){\Large 0}%
\put(211.5,69.5){\Large 0}%
\put(145,43){$p_{i,j-1}$}%
\put(179,43){$p_{i,j}$}%
\put(175,13){$p_{i+1,j}$}%
\put(175,73){$p_{i-1,j}$}%
\put(205,43){$p_{i,j+1}$}%
\end{picture}\\
\caption{Target cells for a pedestrian in the next time
step~\cite{ExtFFCAMod}.}\label{fig1}
\end{center}
\end{figure}
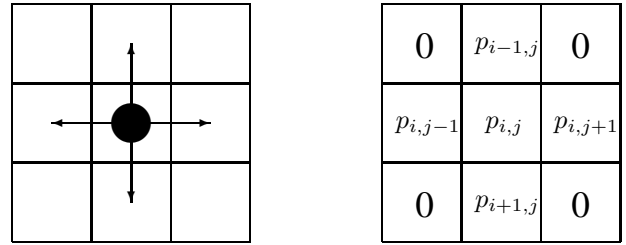

Starting people positions are known. A target point for each
pedestrian is the nearest exit. Each particle can move to one of
four its next-neighbor cells or to stay in present cell (the von
Neumann neighborhood) at each discrete time step $\,t \rightarrow
t+1$ --- fig.~\ref{fig1}; i.e., $v_{max}=1$.

Generally speaking, a direction for each particle at each time
step is random and determined in accordance with a transition
probabilities distribution (and transition rules).

So a main problem is to determine ``right'' transition
probabilities (and transition rules).

\section{Solution}
\subsection{Update rules} A typical scheme
for stochastic CA models is used here. There is step of some
preliminary calculations. Then at each time step transition
probabilities are calculated, and direction is chosen. If there
are more then one candidates to one cell a conflict resolution
procedure is applied, and then a simultaneous transition of all
particles is made.

In our case the {\it preliminary step} includes calculations of
SFF $S$. Each cell $S_{i,j}$ saves shortest discreet distance to
the nearest exit. The unit of such distance is a number of steps.
To calculate the field $S$ (and only here) we admit diagonal
transitions and consider that a vertical and horizontal movement
to the nearest cell has a length of $1$; a length of a diagonal
movement to the nearest cell is $\sqrt 2$. (It's clear that
movement through a corner of walls or collums is forbidden and
around movement is admitted in such cases only.) It is made a
discreet distance more close to continuous one.

Probabilities to move from cell $(i,j)$ to each of four the
nearest cells are calculated in the following way:
\begin{multline}
\label{pij} p_{i-1,j}=\frac{\tilde p_{i-1,j}}{Norm_{i,j}},
p_{i,j+1}=\frac{\tilde p_{i,j+1}}{Norm_{i,j}},\\
p_{i+1,j}=\frac{\tilde p_{i+1,j}}{Norm_{i,j}},
p_{i,j-1}=\frac{\tilde p_{i,j-1}}{Norm_{i,j}},
\end{multline}
\noindent where $Norm_{i,j} = \tilde p_{i-1,j} + \tilde p_{i,j+1}
+ \tilde p_{i+1,j} + \tilde p_{i,j-1}$.

Moreover
\begin{equation}
p_{i-1,j}=0,\quad p_{i,j+1}=0, \quad p_{i+1,j}=0, \quad
p_{i,j-1}=0
\end{equation}
\noindent only if
\begin{equation}
w_{i-1,j}=1,\quad w_{i,j+1}=1, \quad w_{i+1,j}=1, \quad
w_{i,j-1}=1
\end{equation}
\noindent correspondingly.

A probability to stay at present cell isn't calculated directly.
But decision rules are organized in a way that such opportunity
may be realized, and a people patience is reproduced by this
means.

\vspace{3mm}

{\bf  Decisions rules} are the following~\cite{Kirik_etalSCSC07}:
\begin{enumerate}
\item If $Norm_{i,j}=0$ then motion is forbidden,
  otherwise a target cell $(l,m)^{\ast}$ is chosen randomly using  the transition probabilities.
%  \begin{enumerate}
  \item
\begin{enumerate}
  \item If $Norm_{i,j}\neq0$ and  $(1-f_{l,m}^{\ast})=1$ then a target
  cell $(l,m)^{\ast}$ is fixed.
 %%%%%%
  \item If $Norm_{i,j}\neq0$ and  $(1-f_{l,m}^{\ast})=0$ then the cell $(l,m)^{\ast}$ is not available for
  moving and a ``people patience'' can be realized. To do it probabilities of
  the cell $(l,m)^{\ast}$ and all other occupied the nearest neighbors are given to an opportunity not to leave the present
  position. A target cell is randomly chosen again among empty neighbors and the present
position.

%In ~\cite{ExtFFCAMod} particles should move to empty cell at each
%time step. %This fact
% delivers many unrealistic fluctuations of the
%flow.
%Here an
%  opportunity not to move and wait when preferable
%  direction will free appears.
  \end{enumerate}
%%%%%%
  \item Whenever two or more pedestrians  have the same target cell,
  the movement of all involved pedestrians is denied with the
  probability $\mu$, i.e. all pedestrians remain at their
  places~\cite{ExtFFCAMod}. One of
  the candidates moves to the desired cell with the probability $1-\mu$ . A pedestrian that is allowed to
  move is chosen randomly.

% A method for parameter $\mu$ adaptation for
%rectangular room with one exit is described
%in~\cite{Kirik_etalSCSC07}.
%%%%%%
\item Pedestrians that are allowed to move perform their
  motion to the target cell.
%%%%%%
\item Pedestrians that stand in exit cells are removed from the room.

%  \item (about clogging)
\end{enumerate}

These rules are applied to all particles at the same time; i.e.,
parallel update is used.

\subsection{How to calculate probability?}
Mostly in this paper we focus on transition probabilities. In
normal situations people choose their route carefully
(see~\cite{HelbingOV} and reference therein). Pedestrians keep a
certain distance from other people and obstacles. The more hurried
a pedestrian is and the more tight crowd is  the more smaller this
distance is. While moving people follow at least two strategies
--- the shortest path and the shortest time.

In FF models people move to the nearest exit, and their wish to
move there doesn't depend on a current distance to the exit.  From
the probability view point this means that for each particle among
all the nearest neighbor cells a neighbor with the smallest $S$
should have the largest probability. So a main driving force for
each pedestrian is to minimize SFF  $S$ at each time step. But in
this case only a strategy of the shortest path is mainly realized,
and a slight regard to an avoidance of congestions is supposed.
This is not realistic for people movement.

An idea to improve a dynamics in a FF model is to introduce an
environment analyzer in a probability formula. It should decrease
an influence of a short path strategy and increase the possibility
to move to a direction with favorable conditions for a moving.
This will provide  some kind of ``trade off'' between two main
strategies.

In this paper we introduce a revised idea of the environment
analyzer~\cite{Kirik_etalSCSC07} and make an attempt to
mathematically formalize a complex decision making process that
people do choosing their path
--- while moving their strategies may vary: cooperate, coincide and compete
depending on a current position and an environment; i.e.,
depending on a place and time.

At first let us present a probability formula and later we are
discussing it in details. For example,the transition probability
to move from a cell $(i,j)$ to the up neighbor is:
\begin{equation} \label{upprob}
\tilde p_{i-1,j}= A_{i-1,j}^{SFF}
A_{i-1,j}^{people}A_{i-1,j}^{wall}(1-w_{i-1,j}).
\end{equation}

\noindent Here
\begin{itemize}
\item $A_{i-1,j}^{SFF}= \exp \left( k_S \triangle S_{i-1,j} \right)$
      --- the main driven force:
      \begin{enumerate}
             \item $\triangle S_{i-1,j}=S_{i,j}-S_{i-1,j}$;
             \item $k_S \geq 0$ --- a sensitivity parameter (model parameter) that can be
             interpreted as the knowledge of the shortest way to the destination point,
             or as a wish to move to the destination point. $k_S = 0$ means that
             pedestrians
             don't use information from the SFF $S$ and move randomly.
             The higher $k_S$ is the more directed is movement of pedestrians.
       \end{enumerate}
      As far as, SFF depict direct distance from each cell to the
      nearest exit then $\triangle S_{i-1,j}>0$ if cell $(i-1,j)$ is closer
      to exit than current the cell $(i,j)$. $\triangle S_{i-1,j}<0$
      if the current cell is closer. And $\triangle S_{i-1,j}=0$ if
      cells $(i,j)$ and $(i-1,j)$ are equidistant to the exit.

      In contrast to other authors that deal with the FF model
      ~(e.g.,~\cite{ExtFFCAMod},~\cite{HeneinWhite07},
      ~\cite{YanagisawaNishinari},~\cite{ ValidCAModFundDiagr}) and use pure values of the field $S$ in the
      probability formula we propose to use only $\triangle S_{i-1,j}$.
      From a mathematical view point it is the same but computationally
      this trick has a great advantage. Values of SFF may be too high (it
      depends on a size of the space), and $\exp \left( k_S S_{i-1,j} \right)$ is
      uncomputable. This is a significant  restriction of that models. At
      the same time $0\leq\triangle S_{i-1,j} \leq 1$, and problem of
      computing $A_{i-1,j}^{SFF}$ is absent;
%%%%%%%
\item $A_{i-1,j}^{people} = \exp \left( -k_P
      D_{i-1,j}(r^{*}_{i-1,j}) \right)$ --- a factor that takes into account
      a people density in the direction:
      \begin{enumerate}
           \item $r^{\star}_{i-1,j}$ --- a distance  to a the nearest obstacle in this direction
           ($r^{\star}_{i-1,j}\leq r$);
           \item $r>0$ --- a ``visibility'' radius (a model parameter) is a maximal distance
           (number of cells) at which the pedestrian can look through to collect
           information about the density and possible obstacles (but not
           pedestrians);
           \item density $0\leq D_{i-1,j}(r^{*}_{i-1,j})\leq 1$,
                  if all $r^{*}_{i-1,j}$ cells are empty in this direction then $D_{i-1,j}(r^{*}_{i-1,j})=0$,
                  if all $r^{*}_{i-1,j}$ cells are occupied by people in this direction then
                  $D_{i-1,j}(r^{*}_{i-1,j})=1$. We estimate density by using idea
                  of the kernel Rosenblat-Parzen's~\cite{Parzen}%(\cite{Rosenblat},~\cite{Parzen})
                  density estimate, and
                  $$D_{i-1,j}(r^{*}_{i-1,j})=
                  \frac{\sum\limits_{m=1}^{r^{*}_{i-1,j}}\Phi\left(\frac{m}{C(r^{*}_{i-1,j})}
                  \right)f_{i-m,j}}{r^{*}_{i-1,j}},$$
                  were \begin{align}
                       \label{1.14}
                       \Phi(z)&=\begin{cases}
                       \left (0.335-0.067 (z) ^2 \right ) 4.4742, |z| \leq \sqrt{5};\\
                       \qquad \qquad 0;\qquad\qquad \qquad \quad  |z| > \sqrt{5},\end{cases}
                       \end{align}
                  $C(r^{*}_{i-1,j})=\frac{r^{*}_{i-1,j}+1}{\sqrt{5}}$;
            \item $k_P\geq k_S$ --- a people sensitivity parameter (a model parameter)
                  determines an influence of the people density.
                  The higher $k_P$ is  the more pronounced is the strategy of the
                  shortest path.
      \end{enumerate}

%%%%%
\item $A_{i-1,j}^{wall}=\\
      =\exp\left( -k_W(1-\frac{r^{*}_{i-1,j}}{r})\tilde 1(\triangle S_{i-1,j}- \max \triangle S_{i,j})\right)$
      -- a factor that takes into account walls and obstacles:
     \begin{enumerate}
           \item $k_W\geq k_S$ --- a wall sensitivity parameter (a model parameter)
           determines an influence of walls and obstacles;
           \item $\max \triangle S_{i,j} = \\
           \qquad =\max \{\triangle S_{i-1,j},\triangle S_{i,j+1},\triangle S_{i+1,j}, \triangle S_{i,j-1}
           \}$,

           $\tilde 1(\phi)=
                 \begin{cases}
                 0, & \phi < 0, \\
                 1 & \text{otherwise}.
                 \end{cases}$

      An idea of the function $\tilde 1(\triangle S_{i-1,j}- \max \triangle S_{i,j})$ goes from a fact that
      people avoid obstacles only moving towards a destination
      point. But if people take detours (that means not minimizing the SFF)
      approaching to obstacles is not avoiding.
      \end{enumerate}
%%%%
 \item NOTE that only walls and obstacles turn the probability to
      ``zero''.
\end{itemize}

Probabilities to move from cell $(i,j)$ to each of four neighbors
are:
\begin{multline}
\label{up_prob} \tilde p_{i-1,j}= \exp\bigl[ k_S \triangle
S_{i-1,j} - k_P
      D_{i-1,j}(r^{*}_{i-1,j})- \\ - k_W(1-\frac{r^{*}_{i-1,j}}{r})  \tilde 1(\triangle
S_{i-1,j}- \max \triangle S_{i,j})\bigr](1-w_{i-1,j});
\end{multline}
\begin{multline}
\label{right_prob} \tilde p_{i,j+1}= \exp \bigl[ k_S \triangle
S_{i,j+1} - k_P D_{i,j+1}(r^{*}_{i,j+1}) -\\
k_W(1-\frac{r^{*}_{i,j+1}}{r}) \tilde 1(\triangle S_{i,j+1}- \max
\triangle S_{i,j})\bigr](1-w_{i,j+1});
\end{multline}
\begin{multline}
\label{down_prob} \tilde p_{i+1,j}= \exp \bigl[ k_S \triangle
S_{i+1,j} - k_P D_{i+1,j}(r^{*}_{i+1,j})-\\-
k_W(1-\frac{r^{*}_{i+1,j}}{r}) \tilde 1(\triangle S_{i+1,j}- \max
\triangle S_{i,j})\bigr](1-w_{i+1,j});
\end{multline}
\begin{multline}
\label{left_prob} \tilde p_{i,j-1}= \exp \bigl[ k_S \triangle
S_{i,j-1} - k_P D_{i,j-1}(r^{*}_{i,j-1}) -\\
-k_W(1-\frac{r^{*}_{i,j-1}}{r}) \tilde 1(\triangle S_{i,j-1}- \max
\triangle S_{i,j})\bigr](1-w_{i,j-1});
\end{multline}

In \eqref{up_prob}-\eqref{left_prob} a product
$A^{people}A^{wall}$ is the environment analyzer that deals with
people and walls. Parameters $k_P$ and $k_W$ allow to tune
sensitivity of the model to the people density and the approaching
to obstacles correspondingly. And as far as $0\leq\triangle S \leq
1$, $0\leq D(r^{*})\leq 1$ and $0\leq1-\frac{r^{*}}{r}\leq 1$ both
parameters shouldn't be less then $k_S$. The term $A^{wall}$ is
only to avoid obstacles ahead, we will not discuss it here and let
$k_W=k_S$.

The following the shortest path strategy means to take detours
around high density regions if it is possible. The term
$A^{people}$ works as a reduction of the main driving force (that
provides the shortest path strategy), and probability of detours
becomes higher. The higher $k_P\geq k_S$ is the  more pronounced
the shortest time strategy is. Note that the low people density
makes influence of $A^{people}$ small, and the probability of the
shortest path strategy increases for the particle.

\begin{figure}[!t]
\centering
\includegraphics[scale=0.4]{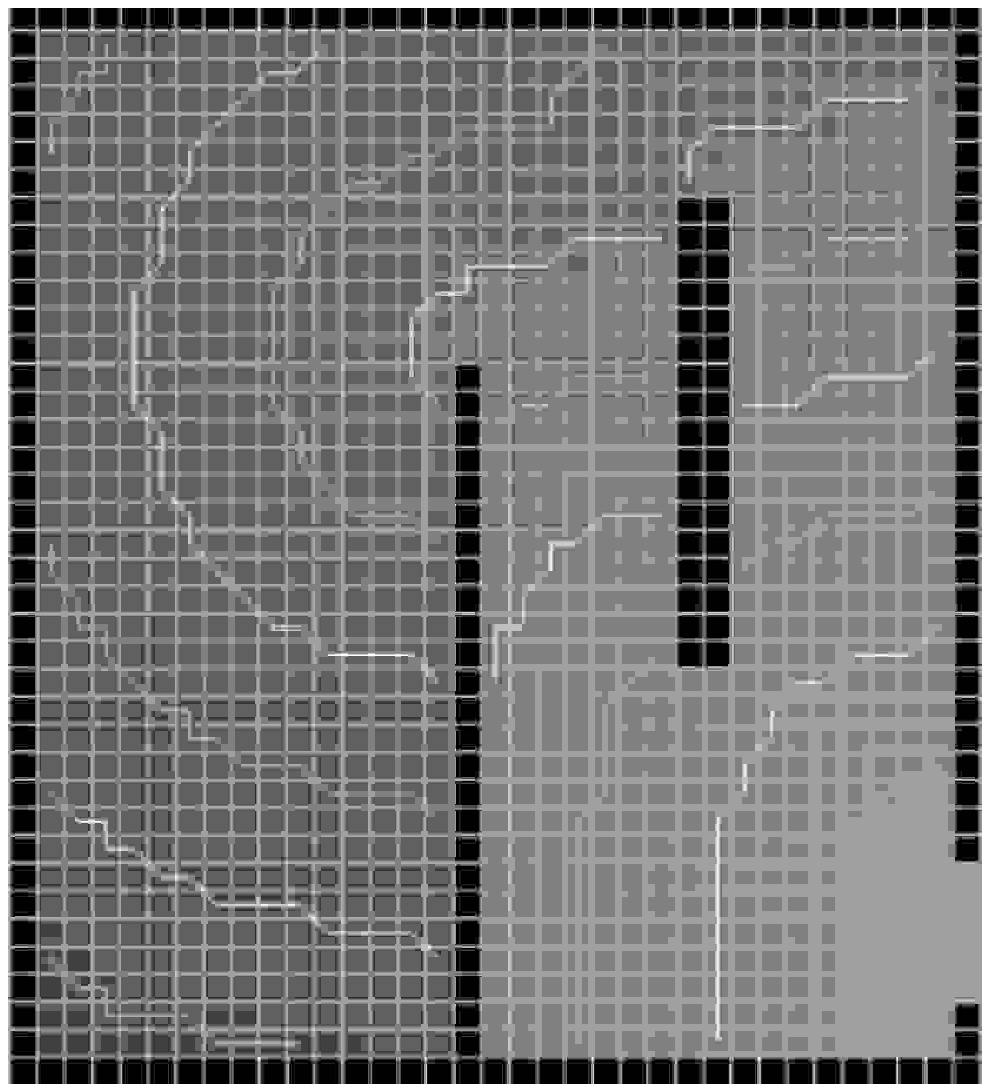}
\includegraphics[scale=0.4]{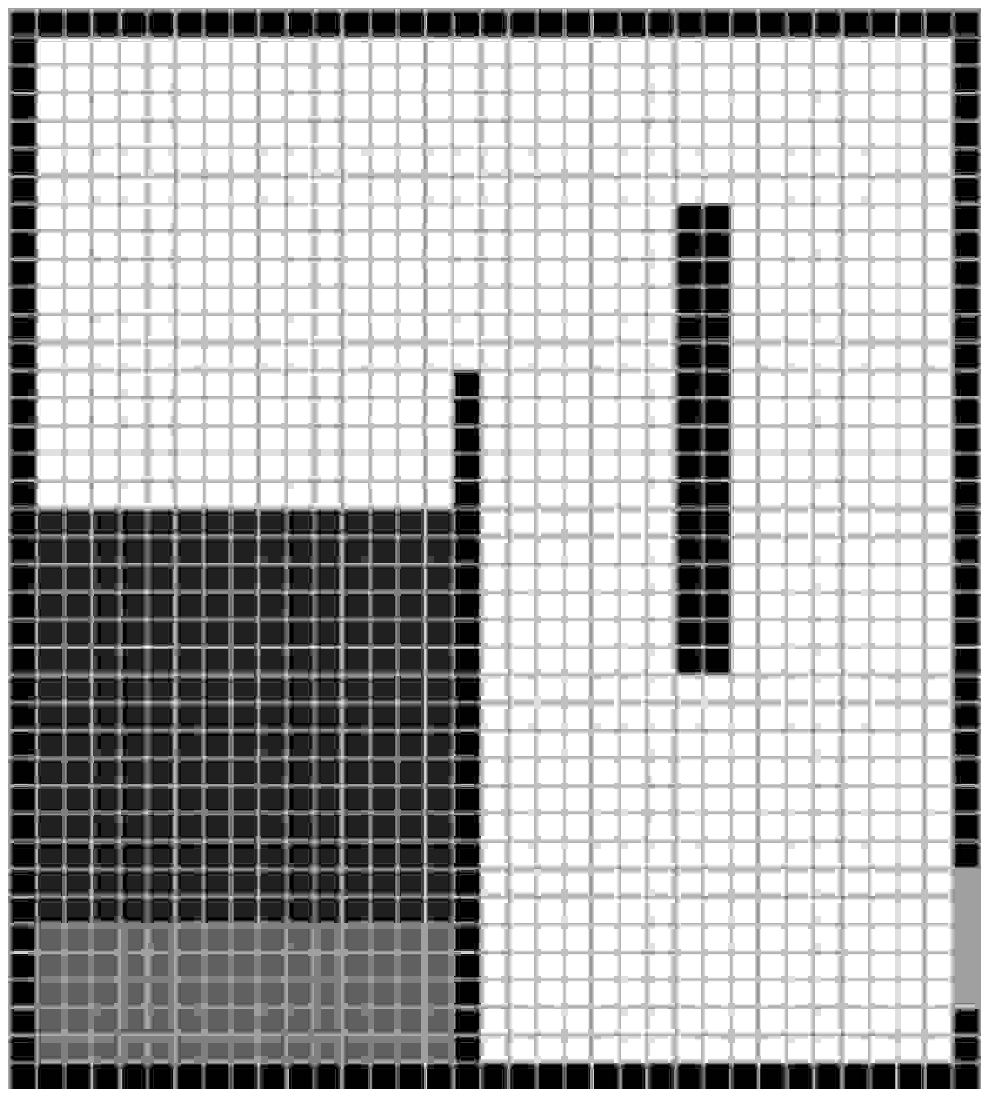}\\
\hspace{0.5cm}\parbox[t]{3cm}{a) F\hspace{-0.4mm}ield $S$.}
\hspace{0.5cm}\parbox[t]{4cm}{b) Initial positions.}\\
\caption{} \label{fig_sim1}
\end{figure}

\section{Simulations}
Here we present some simulation results to demonstrate that our
idea works. We use one space and compare 2 sets of parameters.
Size of space is $14.8m\times 13.2m$ ($37$ cells $\times$ $33$
cells) with one exit ($2.0m$). Recall that the space is sampled
into cells of size $40cm \times 40cm$ which can either be empty or
occupied by one pedestrian only. The static field $S$ is presented
in fig.~2a. In fig.~2b are stating positions of particles. They
move towards  exit with $v=v_{max}=1$.

Here we don't present some quantity results and only demonstrate a
quality difference of the flow dynamics for 2 sets of model
parameters for the model presented.

The first set of parameters is $k_S=k_W=4$, $k_P=6$, $r=10$. The
second set is $k_S=k_W=4$, $k_P=18$, $r=10$. A following moving
condition are reproduced by both sets -- pedestrians know a way to
the exit very well, they want go to the exit (it is determined by
$k_S$), a visibility is good ($r$), attitude to walls is ``loyal''
($k_W=k_S$). Only one parameter varies here, it's $k_P$.

In the first case ($k_P=6$) a prevailing moving strategy is the
shortest path. Fig.~3 presents an evacuation in different moments
for this case.
\begin{figure}%[!t]
\begin{center}
\hspace{0.5cm}\parbox[t]{2cm}{$t=25$}
\hspace{0.5cm}\parbox[t]{2cm}{$t=65$}
\hspace{0.5cm}\parbox[t]{2cm}{$t=135$}\\
\includegraphics[scale=0.25]{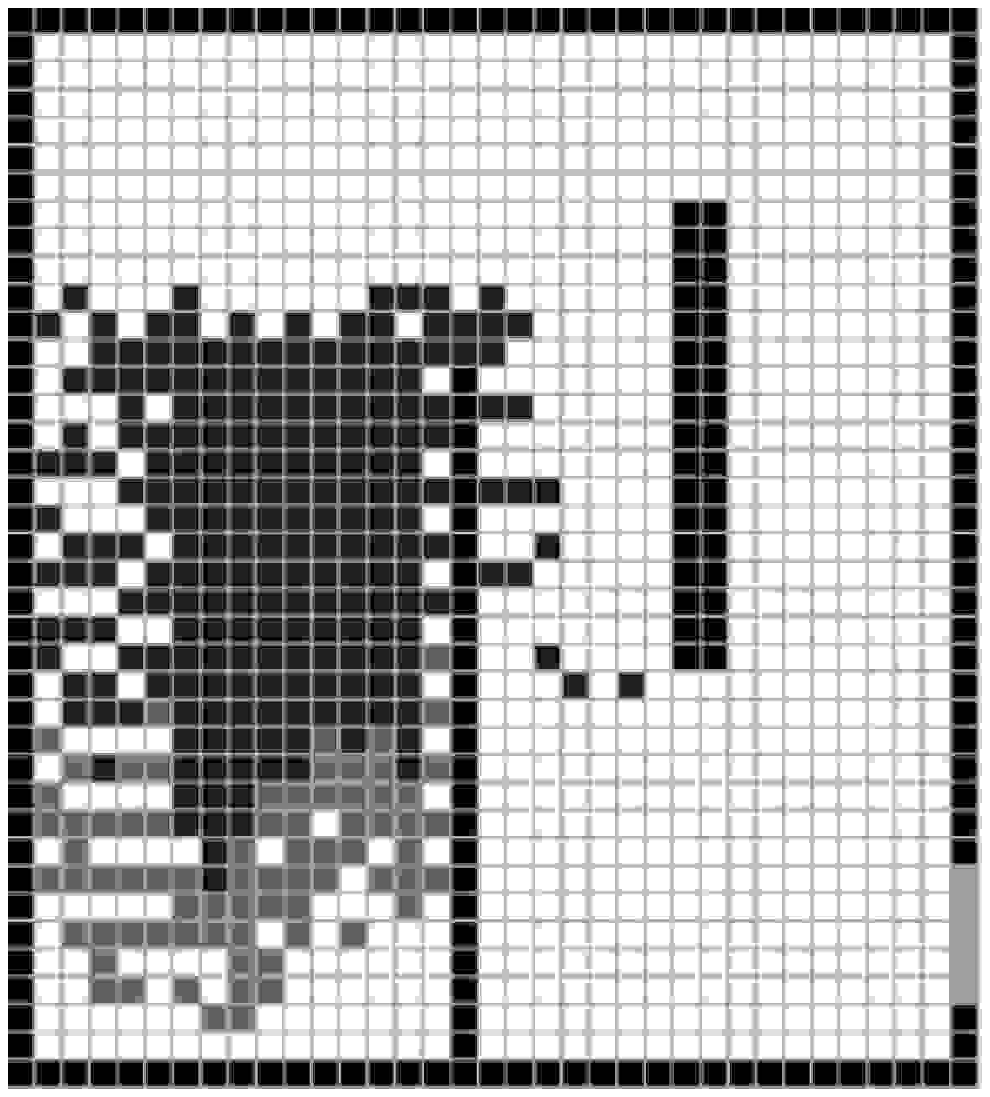}
\includegraphics[scale=0.25]{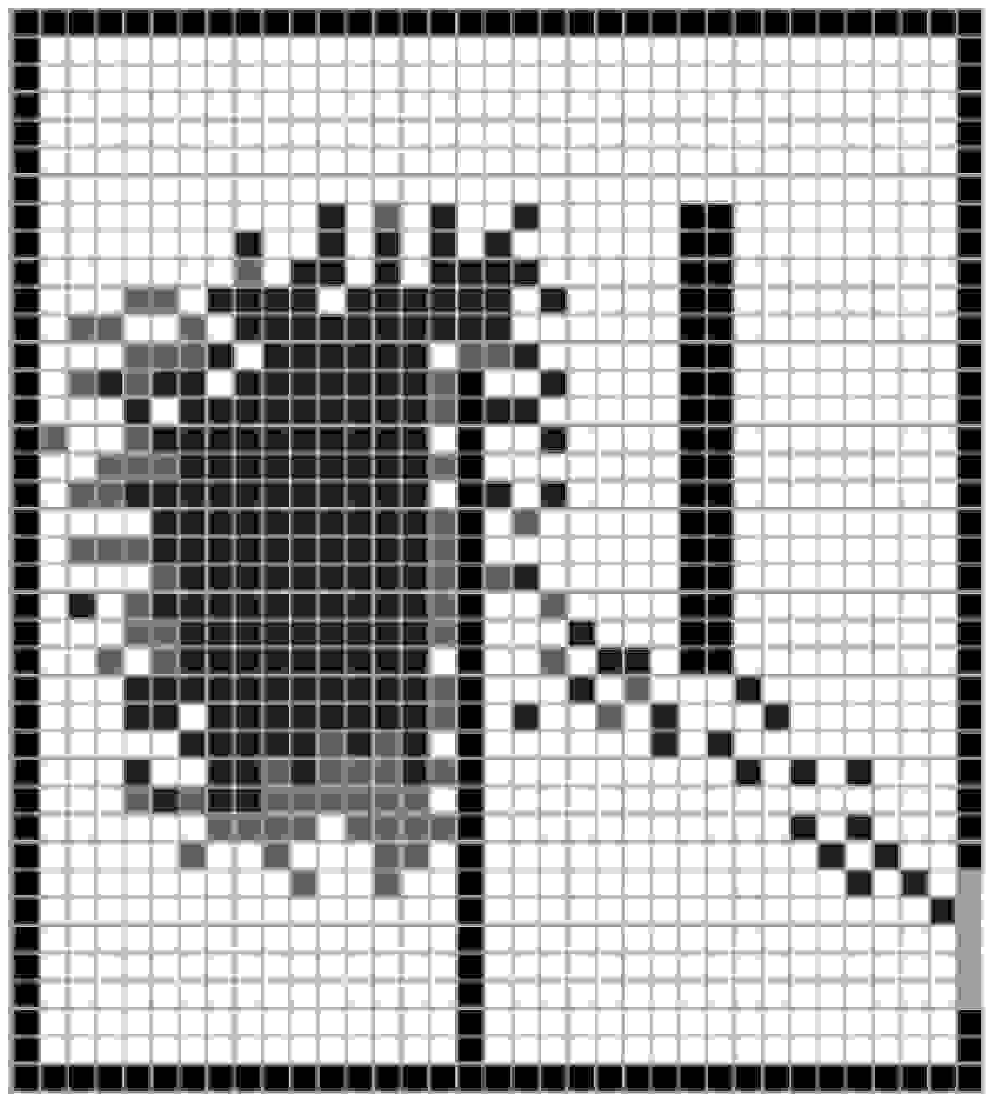}
\includegraphics[scale=0.25]{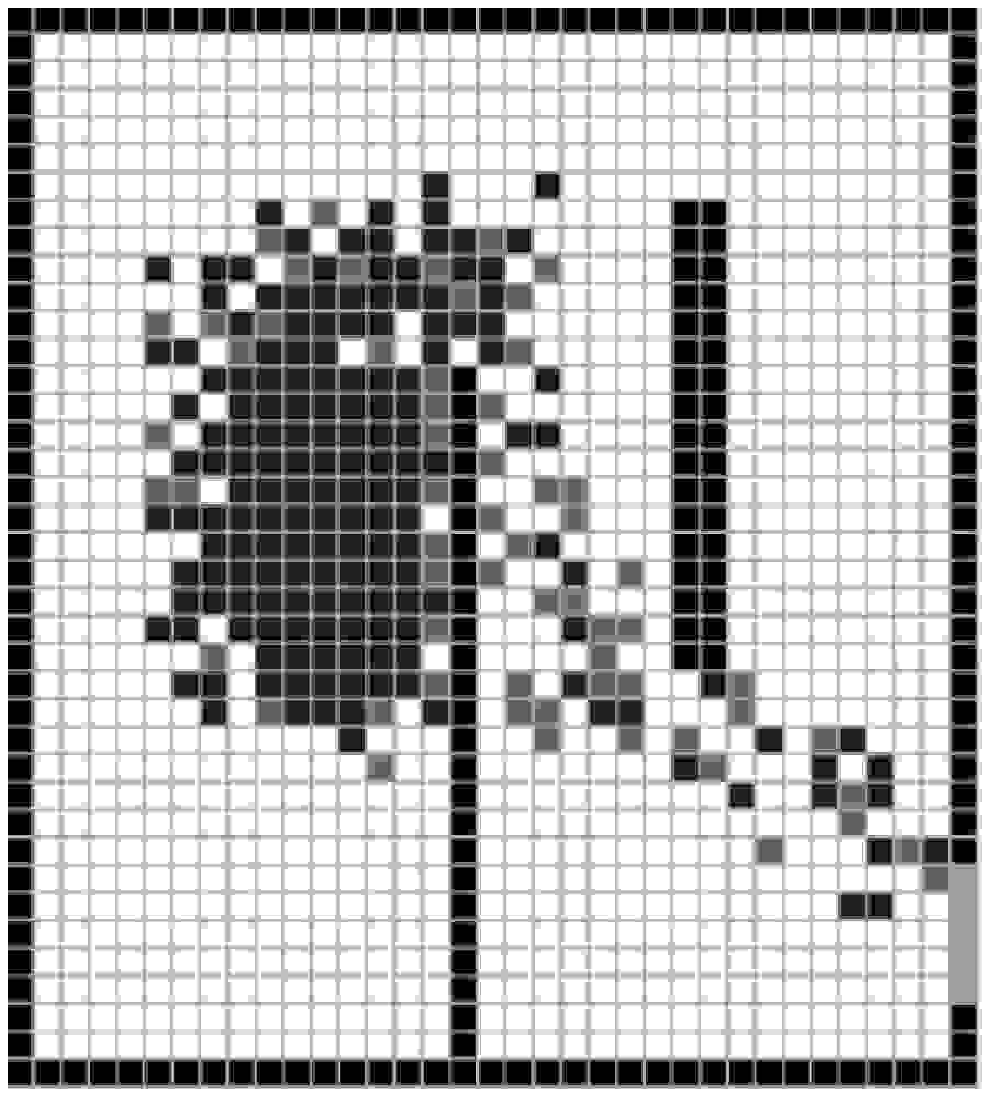}\\
%\hspace{0cm}\parbox[t]{8cm}{a) $k_S=k_W=4$, $r=10$, $k_P=6$}\\[10pt]
%
\hspace{0.5cm}\parbox[t]{2cm}{$t=165$}
\hspace{0.5cm}\parbox[t]{2cm}{$t=180$}
\hspace{0.5cm}\parbox[t]{2cm}{$t=225$}\\
\includegraphics[scale=0.25]{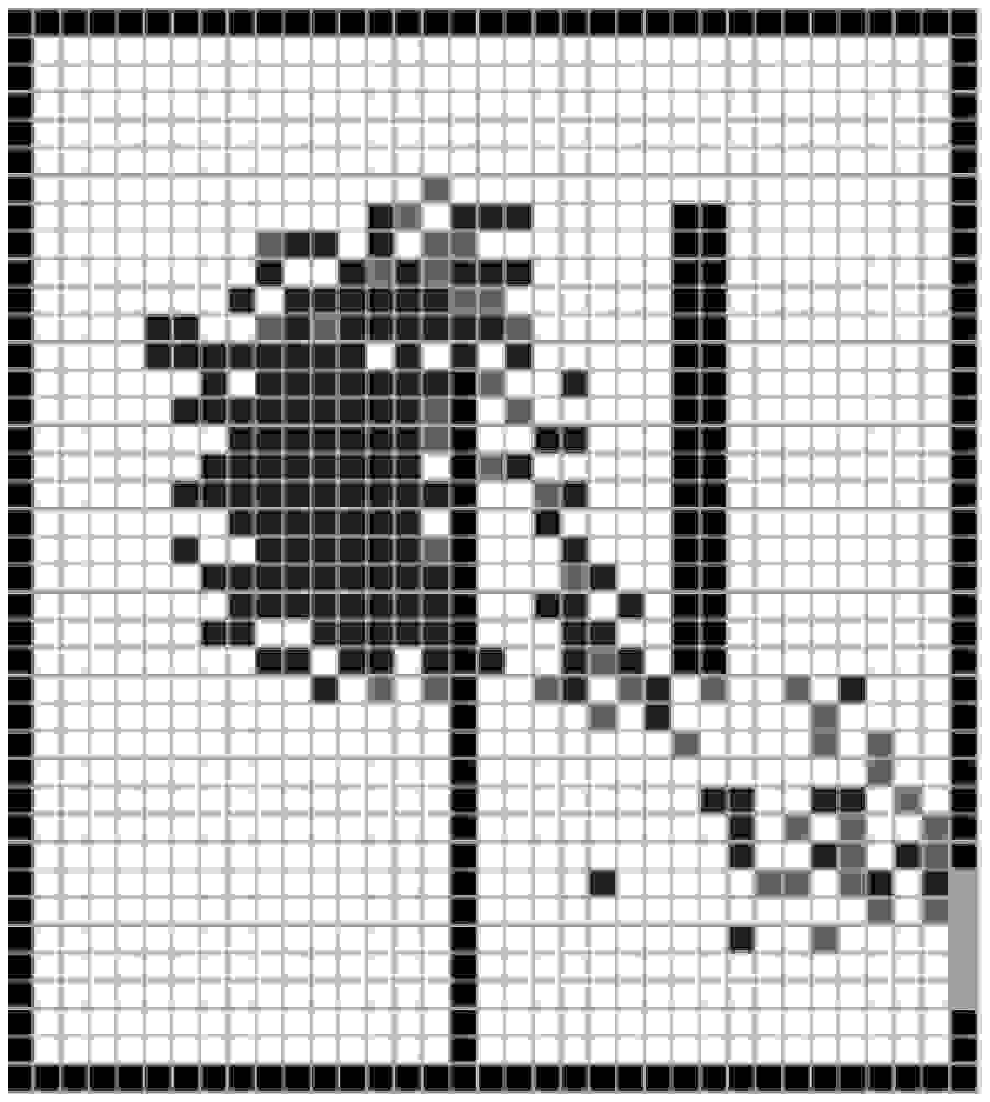}
\includegraphics[scale=0.25]{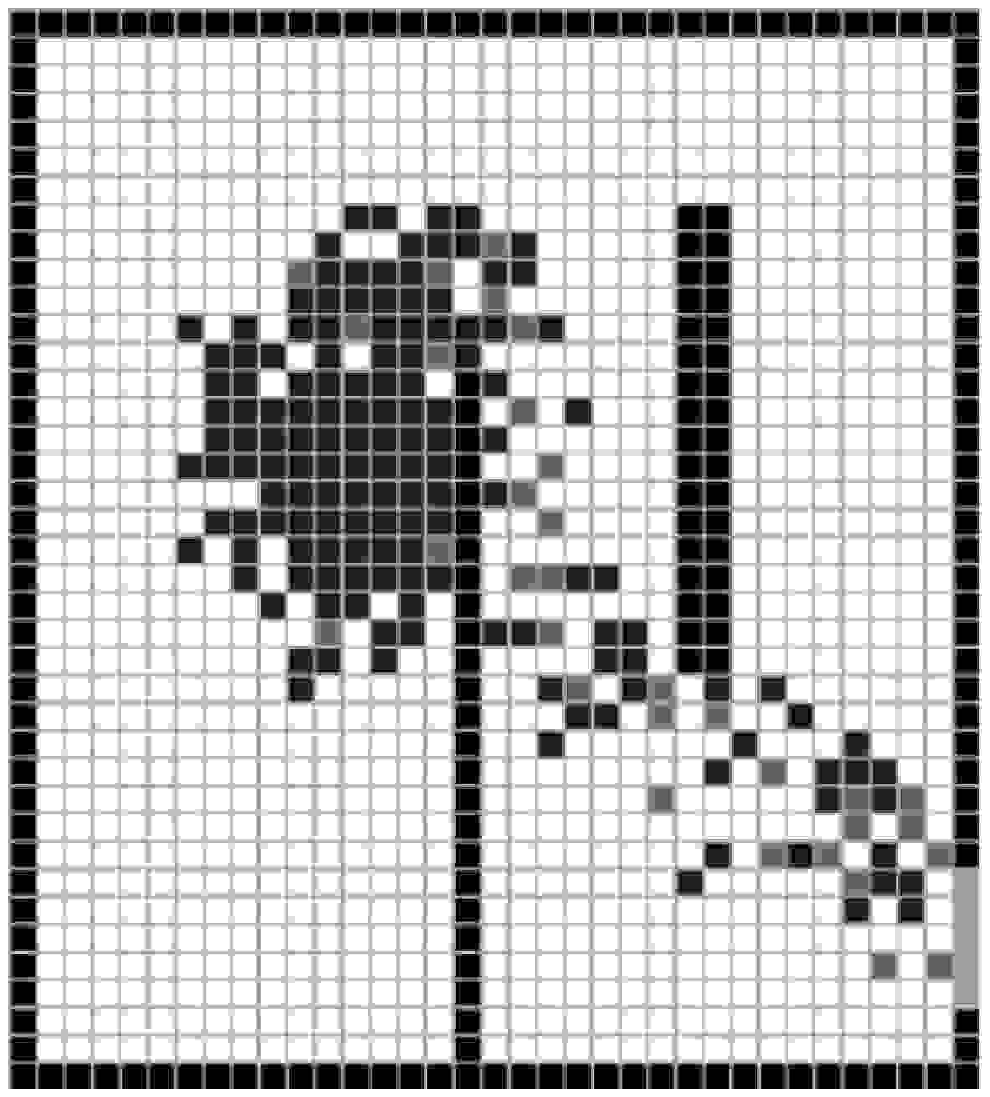}
\includegraphics[scale=0.25]{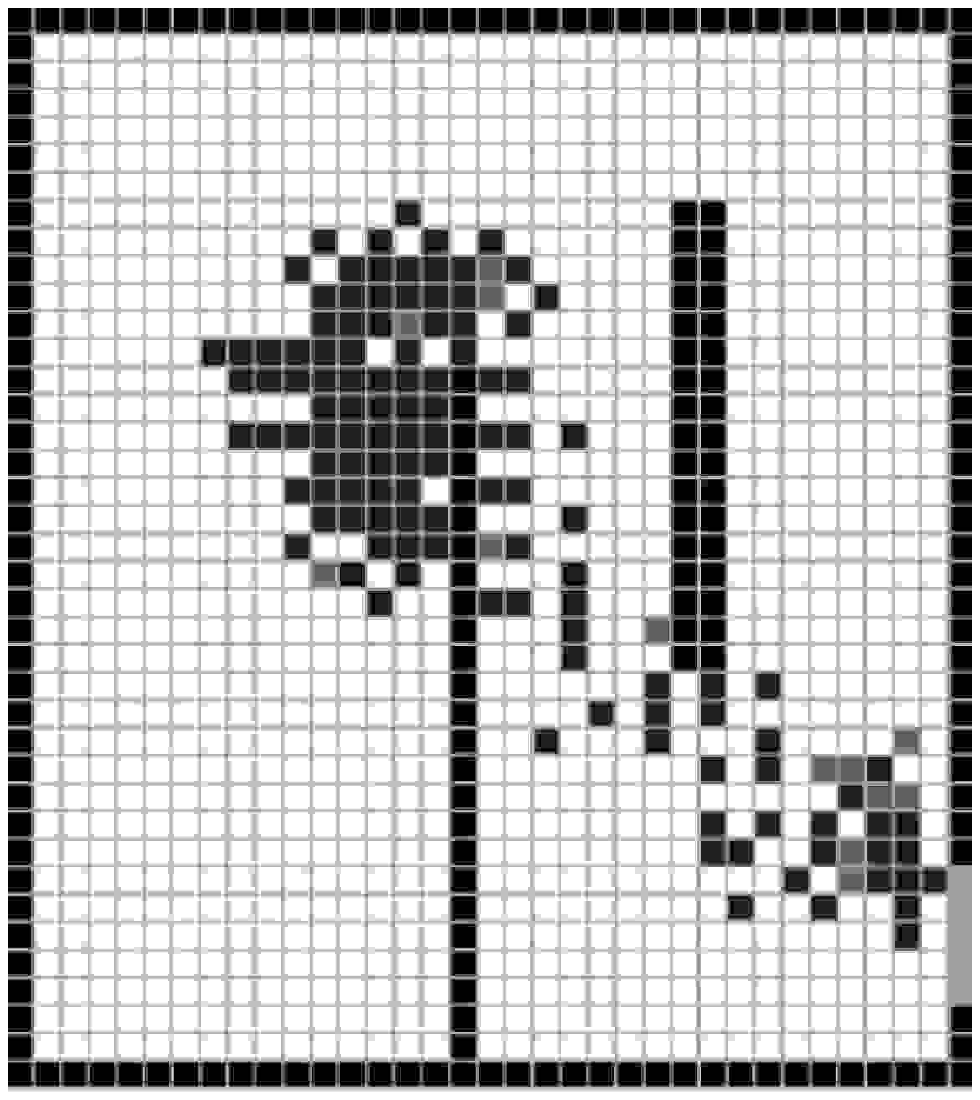}\\
%\hspace{0cm}\parbox[t]{8cm}{b) $k_S=k_W=4$, $r=10$, $k_P=6$}
\caption{Evacuation for 300 people, $k_S=k_W=4$, $r=10$, $k_P=6$.}
\label{300people_ks6}
\end{center}
\end{figure}

The other set of parameters $k_S=k_W=4$, $k_P=18$, $r=10$ (see
fig.~4) allows to realize both strategies depending on conditions.
Recall that the term $A^{people}$ in
\eqref{up_prob}-\eqref{left_prob} works only if the people density
$D(r^{\star})>0$, and it reduces the probability of the shortest
path strategy depending on density's value.

\begin{figure}%[!t]
\begin{center}
\hspace{0.5cm}\parbox[t]{2cm}{$t=25$}
\hspace{0.5cm}\parbox[t]{2cm}{$t=65$}
\hspace{0.5cm}\parbox[t]{2cm}{$t=135$}\\
\includegraphics[scale=0.25]{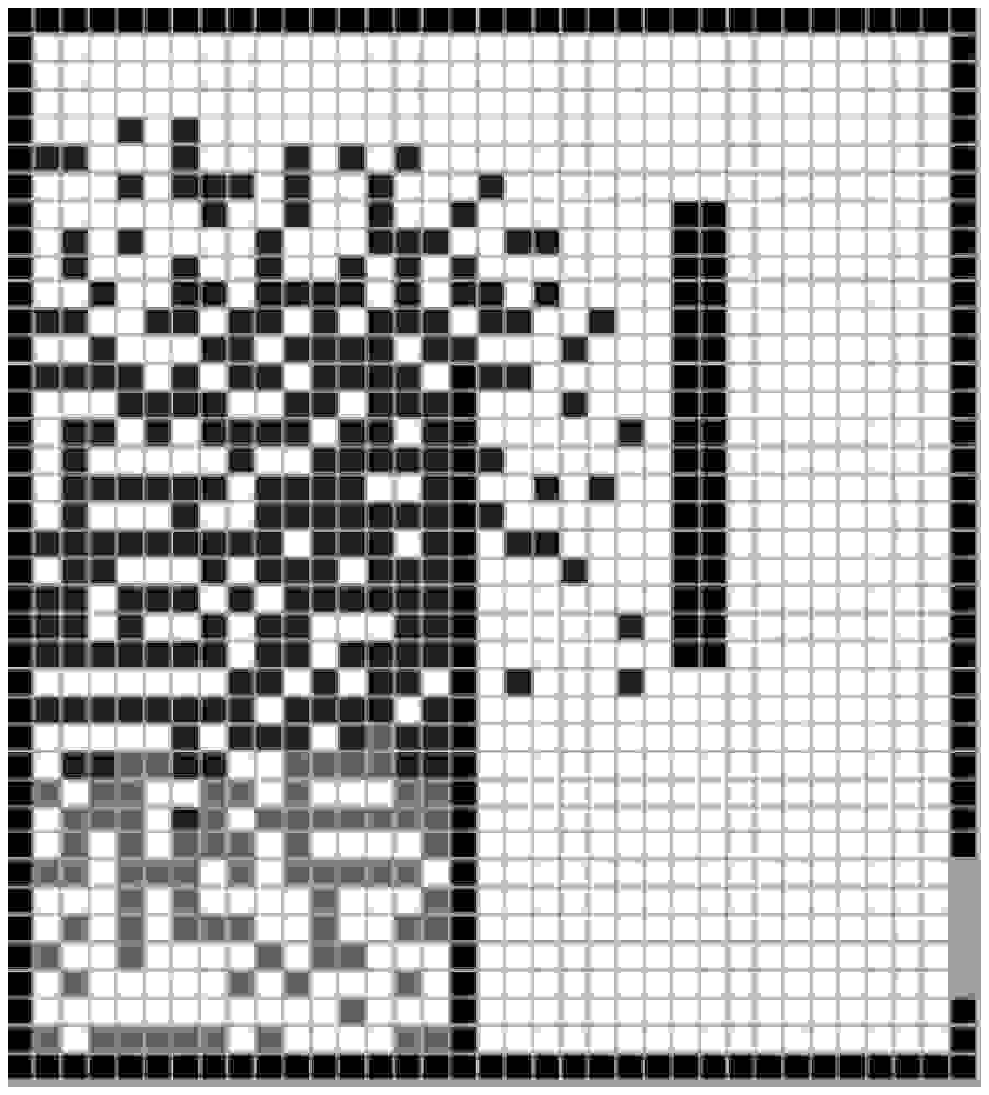}
\includegraphics[scale=0.25]{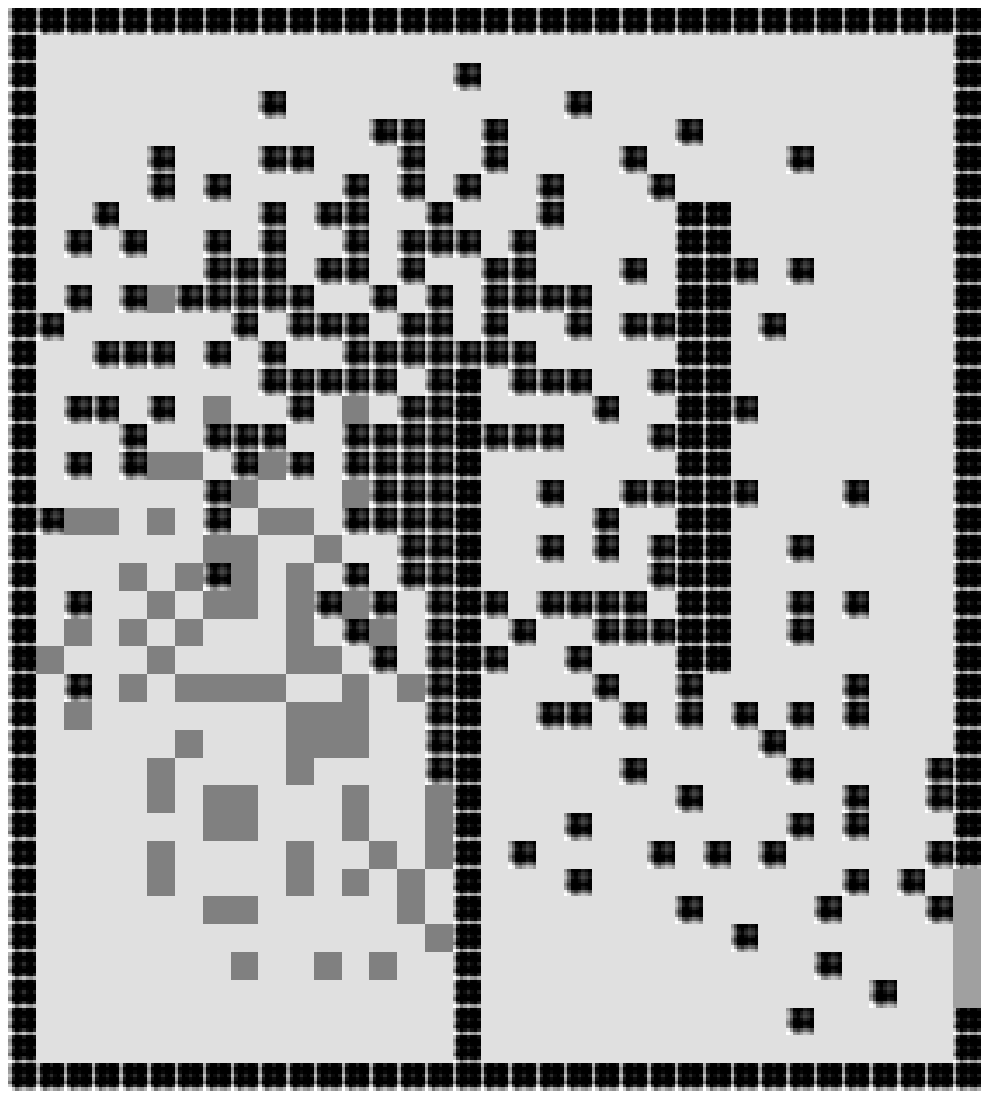}
\includegraphics[scale=0.25]{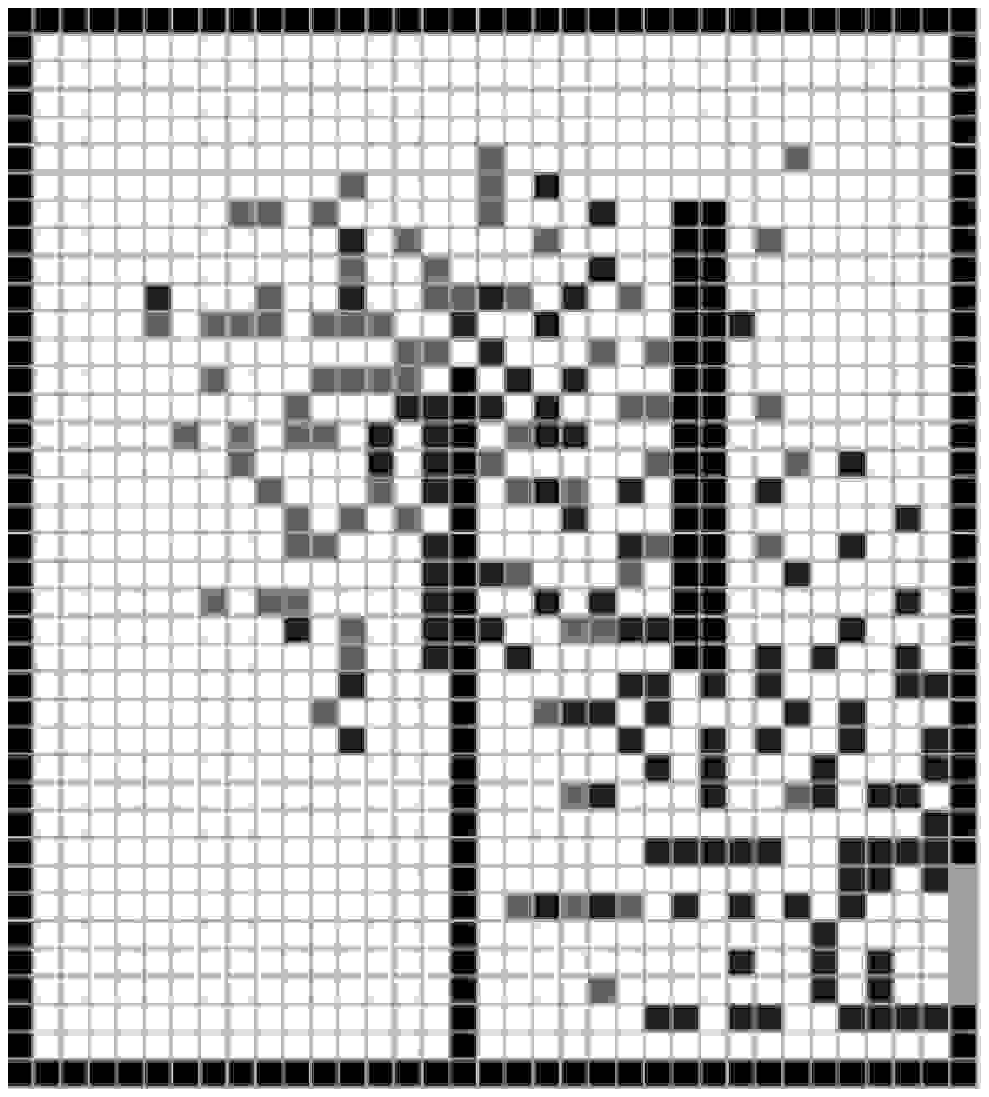}\\
%\hspace{0cm}\parbox[t]{8cm}{a) $k_S=k_W=4$, $r=10$, $k_P=6$}\\[10pt]
%
\hspace{0.5cm}\parbox[t]{2cm}{$t=165$}
\hspace{0.5cm}\parbox[t]{2cm}{$t=180$}
\hspace{0.5cm}\parbox[t]{2cm}{$t=225$}\\
\includegraphics[scale=0.25]{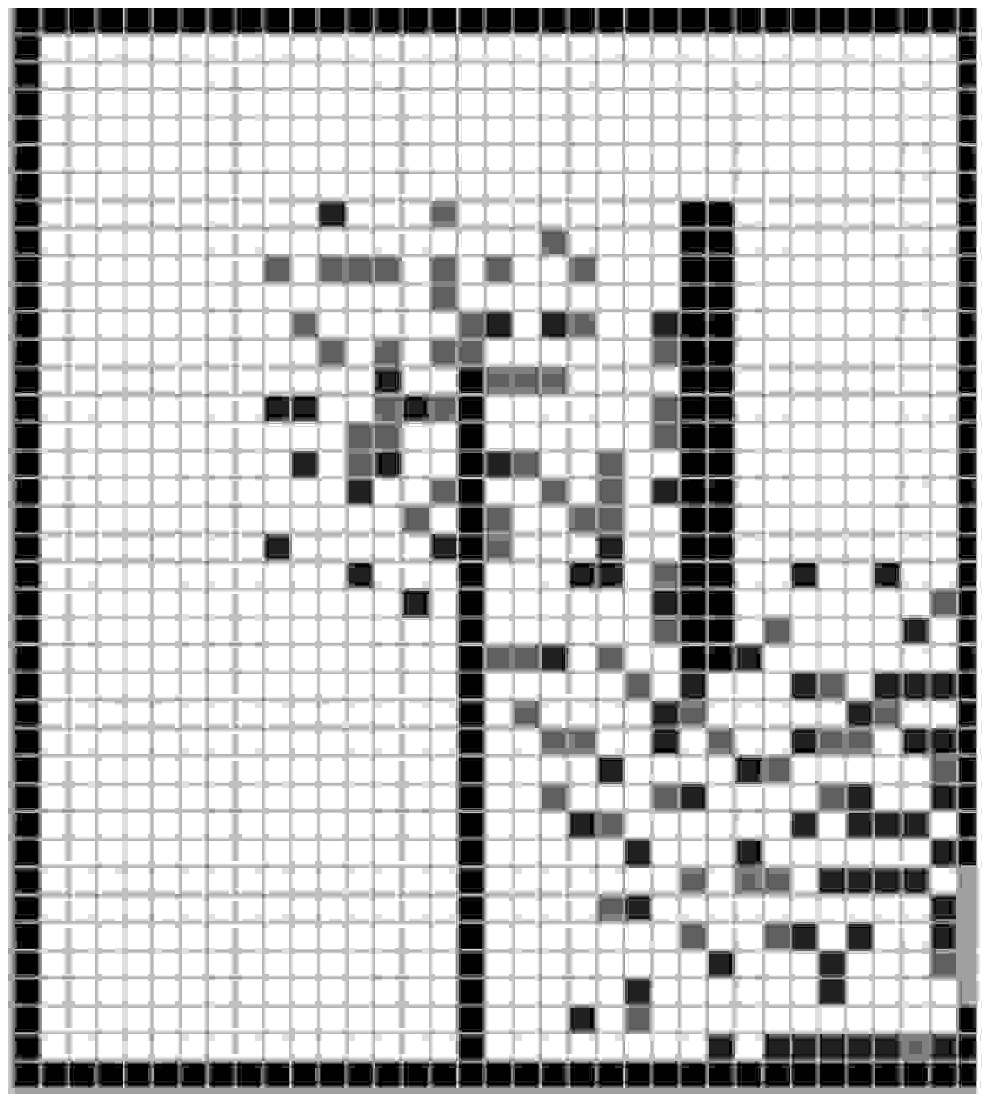}
\includegraphics[scale=0.25]{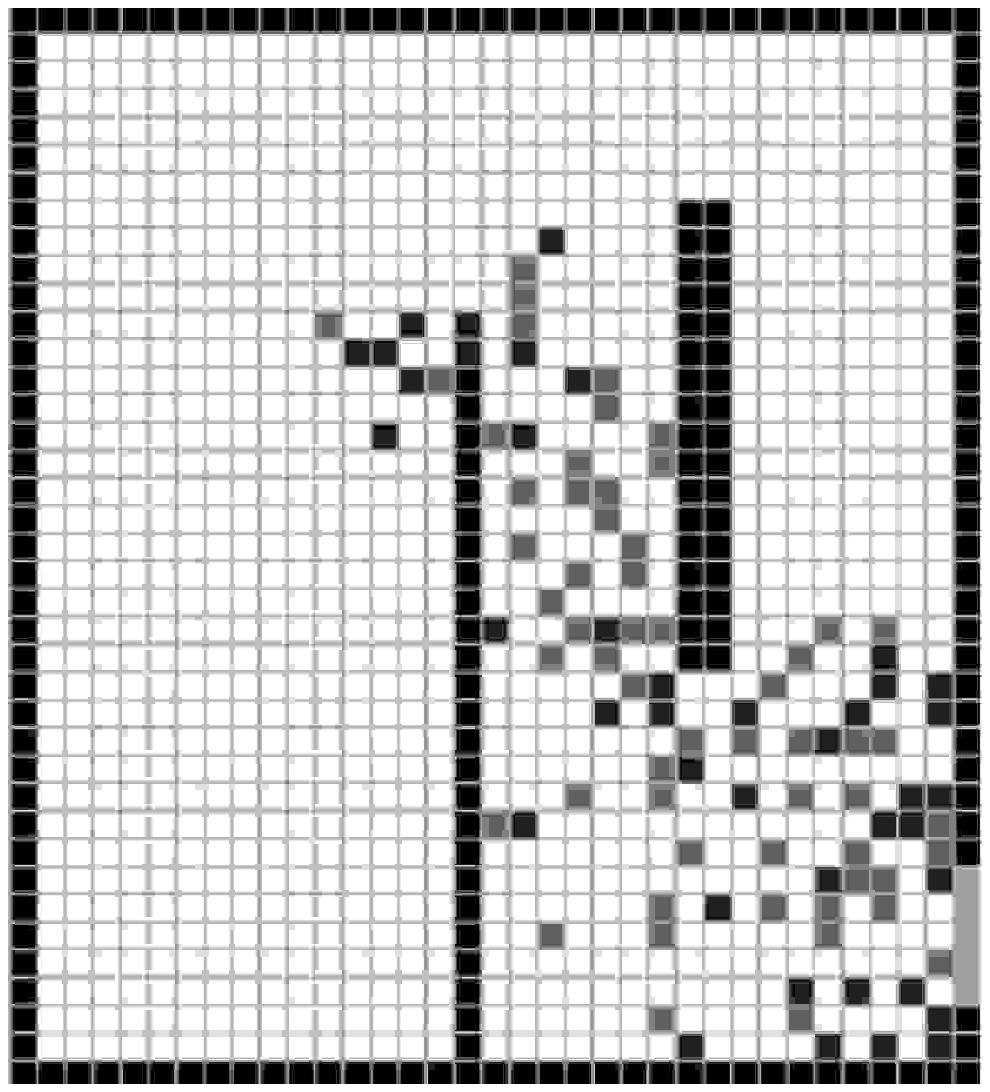}
\includegraphics[scale=0.25]{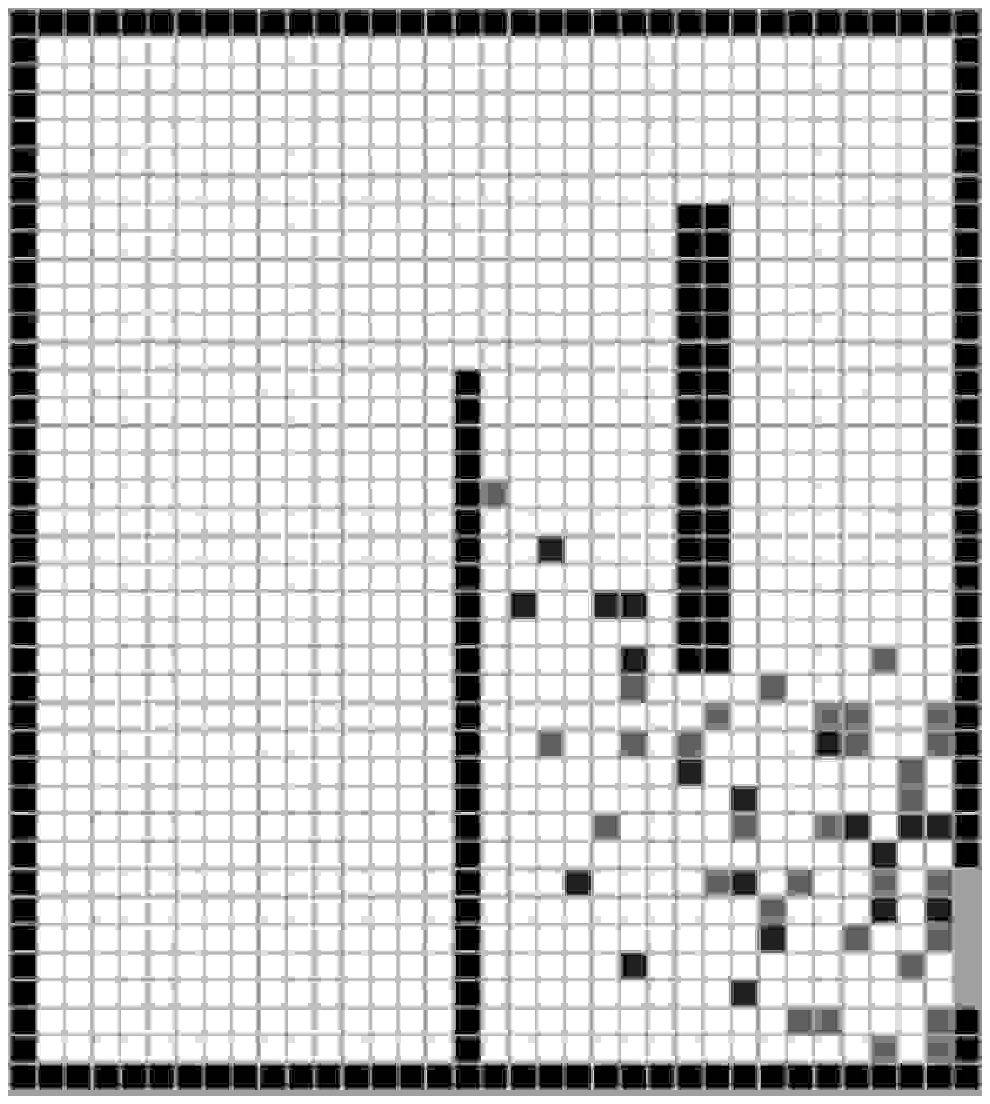}\\
%\hspace{0cm}\parbox[t]{8cm}{b) $k_S=k_W=4$, $r=10$, $k_P=6$}
\caption{Evacuation for 300 people, $k_S=k_W=4$, $r=10$,
$k_P=18$.} \label{300people_ks18}
\end{center}
\end{figure}

\section{Conclusion}
Figures~3-4 show a great difference in the flow dynamics that
obtained  by following only one movement strategy and by ``keeping
in mind'' both strategies at a time. The case of $k_P=18$, i.e.,
when both strategies of the shortest path and the shortest time
are well pronounced, gives a more realistic shape of flow. A model
dynamics proper needs a careful investigation and it is go on. A
necessity of the $k_P$ spatial adaptation is already clear.

% that's all folks
\end{document}